\documentstyle[amsfonts,epsfig,12pt]{article}

\def\mypagenumber{1}

\def\myend{\end{document}}

\normalsize

\newcounter{sxn}

\newcounter{axn}

\date{}

\newdimen\mybaselineskip
\newcommand{\beeq}{\begin{equation}}
\newcommand{\eneq}{\end{equation}}
\newcommand{\be}{\begin{eqnarray}}
\newcommand{\ee}{\end{eqnarray}}
\newcommand{\bpic}{\begin{picture}}
\newcommand{\epic}{\end{picture}}

\def\la{\raise.16ex\hbox{$\langle$} \, }
\def\ra{\, \raise.16ex\hbox{$\rangle$} }

\def\psibar{ \psi \kern-.65em\raise.6em\hbox{$-$} }
\def\mbar{ m \kern-.78em\raise.4em\hbox{$-$}\lower.4em\hbox{} }

\def\n@space{\nulldelimiterspace=0pt \mathsurround=0pt }
\def\huge#1{{\hbox{$\left#1\vbox to 20.5pt{}\right.\n@space$}}}

\def\myskip{\noalign{\kern 8pt}}
\def\myeqspace{\noalign{\kern 10pt}}

\def\boxit#1{$\vcenter{\hrule\hbox{\vrule\kern3pt
    \vbox{\kern3pt\hbox{#1}\kern3pt}\kern3pt\vrule}\hrule}$}
\def\bigbox#1{$\vcenter{\hrule\hbox{\vrule\kern5pt
     \vbox{\kern5pt\hbox{#1}\kern5pt}\kern5pt\vrule}\hrule}$}

\def\ignore#1{{}}

\begin{document}

\bibliographystyle{unsrt}
\footskip 1.0cm

\thispagestyle{empty}
\setcounter{page}{\mypagenumber}

             
\begin{flushright}{UFIFT-HEP-00-28\\}
\end{flushright}
\begin{flushright}{OUTP-00-55-P\\}
\end{flushright}

\vspace{2.5cm}
\begin{center}
{\LARGE \bf {Kaluza-Klein Vortices
}}\\ 

\vspace{2cm}
{\large Vakif K. Onemli$^{a,}$\footnote{e-mail:~ 
onemli@phys.ufl.edu},
\hskip 0.3 cm 
Bayram Tekin$^{b,}$\footnote{e-mail:~
tekin@thphys.ox.ac.uk} }\\
\vspace{.5cm}
$^a${\it Physics Department, University of Florida, Gainesville, 
FL 32611, USA}\\  
\vspace{.5cm}
$^b${\it Theoretical Physics, University of Oxford, 1 Keble Road, Oxford,
OX1 3NP, UK}\\

\end{center}

\vspace*{2.5cm}


\begin{abstract}
\baselineskip=18pt
We study static vortex type solutions of pure gravity for  $D \geq 4+1 $. Non-singular vortex solutions can be
obtained by considering periodic Kaluza-Klein monopoles. 
We also show that away from the center of the vortices the space is described by the gravitational instantons
derived from minimal surfaces. 
\end{abstract}
\vfill


 
\newpage



\normalsize
\baselineskip=22pt plus 1pt minus 1pt
\parindent=25pt

Solitons of gravity theories are extremely important objects since they can be interpreted as nonsingular particles/branes.  
But a theorem by Einstein and Pauli \cite{Einstein} rules out soliton solutions 
of {\it{four}} dimensional (pure) gravity [See also a recent review \cite{Dongen}.] On the other hand theories defined
in more than four dimensions accommodate many interesting soliton solutions. The first example of a pure gravity soliton
is the  Kaluza-Klein monopole
in five dimensions found by Sorkin \cite{Sorkin}, Gross and Perry. \cite{Gross}. The main point of their construction is that
 4 dimensional gravitational instantons solve the static equations of 4+1 dimensional gravity. In order to obtain the 
Dirac monopole 
type particle one uses the Taub-NUT gravitational instanton \cite{Hawking} whose monopole-like structure is well known
[See for example \cite{Eguchi}]. The singularity of the monopole in this construction disappears when one of the spatial 
dimensions is compactified. 
It is amusing to note that  besides  the perfectly smooth `t Hooft- Polyakov monopole  one has an other kind 
of smooth monopole : the Kaluza-Klein monopole. The first one lives in 3+1 dimensions and it is a composite object of finite 
size which is a solution of the equations of a non-Abelian gauge theory with an effective compact $U(1)$ gauge group.
But when one is looking from a distance bigger than the inverse mass of the vector bosons then 
one ``sees'' a Dirac monopole. In the Kaluza-Klein monopole near the origin the spacetime is 4+1 dimensional but if 
one looks from a distance larger then the compactified radius the spacetime becomes 3+1 dimensional and the KK monopole 
looks like exactly the Dirac and the `t Hooft Polyakov monopole.

The story of KK monopole becomes more interesting if one considers it as a solution of low energy string theory or M theory
equations of motion. Townsend \cite{Townsend1, Townsend2} showed that M-Theory compactified on $S^1$ has a one form 
vector potential whose 6 form dual can be coupled to a (magnetic) 6-brane which turns out to have a KK monopole interpretation.
From supergravity point of view this basically means that a 6-brane curves its transverse space to be a Taub-NUT space.

Our aim in this paper is to obtain static vortex type solutions of pure gravity. 
The magnetic field of a KK monopole (or D6 brane in M theory) is spherically symmetric and  spreads to 3 dimensions. 
We would like to find solutions which are gravitational vortices which have magnetic fields trapped in a two dimensional 
spatial slice.
Similar solutions were considered in different contexts \cite{Sanchez, Witten, Vafa}.

In the last part of the paper we relate these vortex solutions to the ``gravitational instantons''
derived from minimal surfaces in  ${\mathbb{R}}^{3}$ \cite{Nutku}.

Kaluza-Klein monopole in M-theory is a spacetime of the form, 
\be
ds^2= -dt^2 + \sum_{i=5}^{10} dy^i\, dy^i + V^{-1}(|\vec{r}|)\,( dx^4 + \vec{\omega}\cdot d\vec{r})^2 + 
V(|\vec{r}|)
d\vec{r}^2.
\ee

So the picture is that we have a $D6$ brane , where the spatial longitudinal coordinates are given by
$i= (5,...10)$ and the transverse coordinates are $( x^4, \vec{r}= \{x^1,x^2, x^3\})$. 
The transverse space is a multi-centered Taub-NUT manifold,
where the potential, $V$ and the gauge field $\omega$
are defined in the following way \cite{Hawking}, 
\be
 \vec{\nabla} \times \vec{\omega}=  \vec{\nabla} V.
\label{potential}
\ee

Apart from $\delta$ function singularities the solution to the above equation $R^3$ is given as
\be
V(|\vec{r}|)= 1+ \sum_{a= 1}^N {4m_a\over |\vec{r} - \vec{r_a}|}
\ee
One can think of the arbitrary parameters, $m_a$ and $r_a$ respectively as the charges and the locations of the KK monopoles
(or D6 branes). The singularities at the locations of the monopoles can be cured as usual 
if one is willing to accept that all the charges are equal, $m_a= m$ and
at the same time $x^4$ becomes compact with a period of $16\pi m$. So M-Theory becomes compactified on 
a circle of radius $8m$.

We search for {\it{vortex}} type soliton solutions 
to pure gravity. For this purpose let us suppress the spatial world-volume coordinates of the D6-brane and
concentrate on the $4+1$ dimensional gravity.
\be
ds^2= -dt^2 +  V^{-1}(|\vec{r}|)\,( dx^4 + \vec{\omega}\cdot d\vec{r})^2 + V(\vec{r})
d\vec{r}^2.
\ee
We know that static and self-dual solutions with 3 dimensional spherical symmetry  correspond to the 
KK monopoles. The explicit form of the metric of course will not be spherically symmetric as one
needs to fix a direction for $\omega$ which would give hedgehog like magnetic field. Namely 
$\vec{\omega}\cdot d\vec{r} = 4m(1-\cos \theta) d\phi$.   
Let us define $y\equiv x^3$  and denote $z = x^1 +ix^2$.
Suppose instead of the above choice for $\omega$; we demand that  $V$ and $\omega$ are independent of $y$ and
we choose $\vec{\omega}(x^1,x^2) \equiv \hat{y} \eta(x^1,x^2)$ and $V(x^1,x^2) \equiv \xi(x^1,x^2)$.
Then the Ricci scalar can be computed as,
\be
R= {1\over 2 \xi^3} \left \{ -\eta_{x^1}^2 + \xi_{x^1}^2 - 2\xi\, (\xi_{{x^1} {x^1}} +\xi_{{x^2} {x^2}})  -
 \eta_{x^2}^2 +\xi_{x^2}^2 \right \}  
\ee
As expected if the equation (\ref{potential}) is satisfied, $\mbox{div} \xi = \mbox{curl} (\hat{y}\eta)$, 
we have a self-dual, Ricci flat space. In two dimensions
this means that we have an analytic function $F(z) = \xi +i\eta$ and Cauchy-Riemann equations are satisfied.
\be
\partial_i \xi= \epsilon_{ij} \partial_j \eta. 
\ee
Therefore $\xi$ and $\eta$ are harmonic,  $\nabla^2 \,\xi = 0$ ,  $\nabla^2 \,\eta = 0$.  
Non-trivial solutions to these equations will necessarily have singularities.
The solution, working in the dimensionless units, is
\be
\eta = \sum_{i=1}^{k}m_i\mbox{Im}\,\mbox{log}(z -z_i)  \hskip 1 cm 
\xi=\sum_{i=1}^{k} m_i\,\mbox{log} |z -z_i|
\ee
$\eta$ is constructed from the angle function on the plane and we allow a $2\pi m_i$ discontinuity as long as
$m_i$ are integers. The multivaluedness of $\eta$ does not yield a multi-valued metric since one can absorb
the discontinuity by a gauge transformation on $x^4$. Thus we have
\be
\eta \rightarrow \eta + 2\pi m_i , \hskip 1 cm  x^4 \rightarrow x^4 -  2\pi m_i y. 
\label{vortex}
\ee

It is clear that this is a vortex solution. The magnetic field, $B_i = \epsilon_{ij}\partial_j \eta$ is 
trapped on the z-plane. Similar configurations arise when one embeds 
four branes to five branes holomorphically in the context of M-theory \cite{Witten}.
Even though there are no exact solutions of a 4-brane-5-brane system, an effective theory
description  shows that the end points of a  4-brane on a 5-brane behave like a vortex.

The problem with our solution is that 
even though the singularity at the origin is milder than
the KK monopole singularity it is not removable by compactifying $x^4$. Therefore    
there are $\delta$-function singularities in the ricci scalar. This singularity is expected since
our solution corresponds to considering many KK monopoles on a line in the y-direction and 
then squeezing them to the z-plane. Whenever all 3 positions of KK monopoles coincide 
there is a singularity. To get rid off this singularity; instead of taking the $V$, $\omega$ to be independent
of y-direction we should impose periodicity in y.    
So we will search for solutions of (\ref{potential}) in $R^2\times S^1$ and  $y \in S^1$. 
A smooth solution  for periodic monopoles located at $(z=z_a, y=y_a)$ is 
\be
V(z, y) = V_0 - {1\over 2} \sum_a \sum_{p= -\infty}^\infty m_a
\left[ { 1\over \sqrt{ |z- z_a|^2 + (y- y_a - 2\pi p)^2}} - {1\over 2\pi |p|} \right ]
\label{summation}
\ee  
The second term in the summation ensures the regularity of the solution at $(z=z_a, y=y_a)$. $m_a$ are
integer charges. For $p=0$, the second term is omited in the sum.
Let us concentrate on a single  periodic monopole with one unit of charge, $m =1$,  
and located at $(z_a=0, y_a=0)$.
Using the Poisson resummation  \cite{Sanchez,Vafa,Gradshtein} one can write this as
\be
 V(z, y) =  1+  {\log |z|^2 \over 4\pi} - {1\over \pi} \sum_{p =1}^{\infty} \cos {p y}K_0 (p|z| )  
\label{solution}
\ee
where $K_0$ is the modified Bessel function. We have normalized the constant $V_0$ to be 
$V_0=1 + {{\log(4\pi) -C}\over 2\pi}$, where $C$ is the Catalan's constant. 
This formula appears in \cite{Vafa} in the context of 
D-Instanton corrections to the {\it{hypermultiplet moduli space}} of type IIA string theory (or M-theory)
compactifications near a conifold singularity. The classical singularity $z=0$ corresponds to a
vanishing 3-cycle and it is resolved in quantum theory by wrapping infinitely many euclidean D2 instantons
around $z=0$. The contributions of the D2 instantons are given by the third term in (\ref{solution}) which
is exponentially suppressed at large $z$.

In this paper our motivation is different as stated above. We consider the solution (\ref{solution}) as
a static vortex solution to $D\geq 4+1$ pure gravity theories.
The singularity at $z=0$ is cured since KK monopoles do not coincide. The large $z$ behavior of the solution
is still given by  the vortex type solution given in (\ref{vortex}). Therefore the magnetic fields read
\be
&&B_{z}= 2\partial_{\bar{z}} V(z,y) = {1\over 2\pi {\bar{z}}} +{1\over \pi} e^{i\arg z}
\sum_p p \cos { p y}K_1 ( p |z| ) , \\ 
&&B_{\bar{z}}= 2\partial_z V(z,y) = {1\over 2\pi z} +{1\over \pi} e^{-i\arg z}
\sum_p p \cos {p y}K_1 ( p |z| ) , \\ 
&&B_y=  \partial_{y} V(z,y) = {1\over \pi}
\sum_p p \sin {p y}K_0 ( p |z| ) . 
\ee
It is clear from the form of magnetic field that at large $z$ only the first terms of $B_{z}$ and  $B_{\bar{z}}$
 survive since the Bessel functions decay exponentially and one has a
hedgehog-like magnetic field in 2 dimensions. As expected it is given by $B = 1/{2\pi |z|}$ with a unit flux.
By construction for small $z$ there is a periodic magnetic field in the $y$ direction which exponetially decay
as $z$ is increased. Multi Kaluza-Klein vortex solutions follow in a similar manner. For this case there is
an other summation over the positions of the vortices as indicated in Eqn. (\ref{summation}).

Next we point out that the large $z$ behavior of these vortices coincide with the ``minimal surface''
solution of the gravitational instantons \cite{Nutku, Aliev}.~\footnote{A similar construction 
exits for gauge theory instantons. In \cite{Comtet, Tekin} it was shown that in ${\mathbb{R}}^{4}$ 
multi-instantons on a line are represented by minimal surface in   ${\mathbb{R}}^{2,1}$. }
Since these minimal surfaces  provide approximate description of the 
gravitational vortices we will sketch their properties below.

The claim is that for every minimal surface in ${\mathbb{R}}^{3}$, there is a gravitational instanton in
four dimensions \cite{Nutku}. As we will see this claim needs some refinement.
One starts with the following ansatz,
\be
ds^2 = {1\over \sqrt{ \mbox{det}g(x^1,x^2)}}\Bigg ( g_{ij}(x^1,x^2) dx^i dx^j + g_{ij}(x^1,x^2) dy^i dy^j \Bigg)
\ee 
We denote  $y^1= y$, $y^2 = x^4$. The Einstein's equations reduce to the minimal surface equation as
\be
f_{x^1\,x^1}\, (1+f_{x^2}^2) -2 f_{x^1}\, f_{x^2}\, f_{{x^1}\,{x^2}} + f_{{x^2}\,{x^2}}\, (1+f_{x^1}^2) =0
\label{minimalsurface}
\ee
$g_{ij}$ is the metric on the minimal surface and  $\mbox{det}g= 1 + f_{x^1}^2 +f_{x^2}^2$. 
A general conformal immersion solution to the minimal surface equation is given by the Weierstrass 
representation.
\be
f= {\mbox{Re}}\,\Bigg( \int (1- h^2, i(1+h^2), 2h)\zeta \Bigg):\Sigma \rightarrow   {\mathbb{R}}^{3} 
\label{Weierstrass}
\ee 
Where  $h$ is a holomorphic function and $\zeta$ is a holomorphic 1-form. 
At first sight this solution seems to 
generate many gravitational instantons, like {\it{helicoid}} defined by $f(x^1,x^2) = \arctan(x^2/x^1)$,
{\it{catenoid}} given as   $f(x^1,x^2)= \cosh^{-1}\sqrt{(x^1)^2 +(x^2)^2} $ or {\it{Scherk surface}} 
 $f(x^1,x^2)= \log{(\cos x^2/\cos x^1})$, etc. But an important theorem by Bernstein states that {\it{there are no
non-trivial solutions of (\ref{minimalsurface}) which are defined on the whole $(x^1, x^2)$ plane}}~\footnote{For a 
proof of the  theorem we refer the reader to the excellent book of Osserman \cite{Osserman}.} 
As it is also clear from the above examples the solutions are singular at some points.
These singularities are not removable therefore one should be careful in calling these solutions
gravitational instantons which are supposed to be complete non-singular solutions.

By a choice of coordinates it was shown that all these solutions 
take the Gibbons-Hawking form \cite{Aliev}.
\be
ds^2= \xi(x^1,x^2) d\vec{r}\cdot d\vec{r}  + {1\over \xi(x^1,x^2)}\, ( dx^4 + \eta(x^1,x^2) dy)^2 
\ee  
In these coordinates self-duality equations reduce to $\partial_i \xi= \epsilon_{ij} \partial_j \eta$,
The singularity dictated by Bernstein's theorem is apparent here because all the non-trivial
solutions of the Laplace equation on the 2 dimensional plane has singularities.
This metric describes the asymptotic behavior of the vortices we have found.

One important point to observe is that
the minimal surface structure of the solution disappears in the Gibbons-Hawking form. 
In fact there is no reference to the ``original''  minimal surface (\ref{Weierstrass}) embedded 
in  ${\mathbb{R}}^{3}$. But in these new coordinates a minimal surface in   ${\mathbb{R}}^{4}$ (instead of
 ${\mathbb{R}}^{3}$) can easily be recognized. The reasoning is as follows. 
As stated above Einstein equations tell us that
we have an analytic function $F(z) = \xi +i\eta$. The graph of an analytic function is always a
minimal surface in four real dimensions \cite{Osserman}. 

In conclusion we have demonstrated that there are static vortex solutions in higher dimensional pure 
gravity theories. The easiest way to construct these solutions is to sum up periodic Kaluza-Klein monopoles.
These vortices appear in a number of different physical contexts as in the case of holomorphically embedding
M4-branes to M5 branes in M-theory \cite{Witten} or in the quantum hypermultiplet moduli space of Type IIA string theory
\cite{Vafa}. Moreover we also pointed out that away from the center of the vortex the space is described by the 
metrics (``gravitational instantons'') derived from minimal surfaces.

V.\ K.\ O. would like to thank Pierre Sikivie and Richard P. Woodard for encouragement and support. V.\ K.\ O. is
supported by DOE Grant DE-FG02-97ER41029.
B.T. would like to thank Ian Kogan and Alex Kovner for discussions.
The research of B.\ T.\ is supported by  PPARC Grant PPA/G/O/1998/00567.

\vskip 1cm


\myend
\begin{thebibliography}{99}

\bibitem{Einstein}
A. Einstein and W. Pauli, ``On the non-existence of regular stationary solutions of 
relativistic field equations'', Ann. Math. {\bf 44} (1943) 131.   

\bibitem{Dongen}
J.van Dongen, `` Einstein and the Kaluza-Klein particle'', gr-qc/0009087.

\bibitem{Sorkin}
R. ~D.~Sorkin,
``Kaluza-Klein monopole,''
Phys.\ Rev.\ Lett.\  {\bf 51}, 87 (1983).

\bibitem{Gross}
D.~J.~Gross and M.~J.~Perry,
``Magnetic monopoles in Kaluza-Klein theories'', 
Nucl.\ Phys.\  {\bf B226}, 29 (1983).

\bibitem{Hawking} 
G. W. Gibbons,  S. W. Hawking, ``Gravitational multi-instantons,''
 Phys. Lett. {\bf 78B} {1978}  430

\bibitem{Eguchi} 
T. Eguchi, A. J. Hanson, ``Self-dual solutions to Euclidean gravity'',
  Annals of Phys. {\bf 120} {1979} 82

\bibitem{Townsend1}
P.~K.~Townsend,
``The eleven-dimensional supermembrane revisited,''
Phys.\ Lett.\  {\bf B350}, 184 (1995)
hep-th/9501068.

\bibitem{Townsend2}
P.~K.~Townsend,
``M-theory from its superalgebra,''
hep-th/9712004.

\bibitem{Witten}
E. Witten  ``Solutions of four-dimensional field theories via M-theory,''
Nucl.\ Phys.\  {\bf B500}, 3 (1997)
[hep-th/9703166].

\bibitem{Sanchez}
N. Sanchez, ``Gravitational calorons'' , Phys. lett. {\bf B125} (1983) 403


\bibitem{Vafa}
H. Ooguri, C. Vafa `` Summing up Dirichlet instantons'' 
 Phys. Rev. Lett {\bf 77} {1996} 3296

\bibitem{Nutku}
Y. Nutku,``Gravitational instantons and minimal surfaces,''
 Phys. Rev. Lett {\bf 77} {1996} 4702

\bibitem{Gradshtein}
I. S. Gradshtein, I. M. Ryzhik, ``Table of integrals, series and products''
Academic Press 1980


\bibitem{Aliev}
A. N. Aliev, J. Kalayci, Y. Nutku,`General minimal surface solution for gravitational instantons,''
  Phys. Rev {\bf D56} {1997} 1332


\bibitem{Osserman}
R.\ Osserman, ``A Survey of minimal surfaces''  
Dover Publications, Inc. New York, 1986


\bibitem{Comtet}
A. Comtet ``Instantons and minimal surfaces'' Phys. Rev {\bf D18} (1978) 3890

\bibitem{Tekin}
B. Tekin ``Multi-instantons in  ${\mathbb{R}}^{4}$ and minimal surfaces in  ${\mathbb{R}}^{2,1}$.
JHEP {\bf 08} (2000) 049

\end{thebibliography}
